\documentstyle[aps,epsf,preprint]{revtex}

\begin{document}

\draft


\title{Hyperon properties in finite nuclei using 
       realistic $YN$ interactions}

\author{I. Vida\~na, A.\ Polls, A.\ Ramos}

\address{Departament d'Estructura i Constituents de la Mat\`eria,
         Universitat de Barcelona, E-08028 Barcelona, Spain}

\author{M.\ Hjorth-Jensen}

\address{ Nordita, Blegdamsvej 17, DK-2100 K\o benhavn \O, Denmark}

\maketitle
\bigskip

\begin{abstract}
Single-particle energies of $\Lambda$ and $\Sigma$ hyperons
in several nuclei are obtained from the relevant
self--energies.
The latter are constructed within the framework of a perturbative
many-body approach  
employing present realistic hyperon-nucleon interactions
such as the models of the J\"{u}lich and Nijmegen groups.
The effects of the non-locality and
energy-dependence of the self--energy on the bound states are
investigated.
It is also shown that, although the single-particle hyperon energies are well
reproduced by local Woods-Saxon hyperon-nucleus potentials, the wave functions
from the non-local self--energy are far more extended. Implications of
this
behavior on the mesonic weak decay of $\Lambda$ hypernuclei are discussed.
\end{abstract}

\bigskip
\bigskip
\newpage


\section{Introduction}

Hypernuclei are bound systems of neutrons, protons and one or more strange 
baryons, such as the $\Lambda$ or $\Sigma$ hyperons. Understanding the
behavior of hypernuclei (how they are produced, their spectroscopy and
decay mechanisms) has been the subject of intense investigations
during the last decades, see e.g., Refs.\
\cite{gal77,povh78,bando85,dover89,oset90,cohen90,bando90,gibson95,akaishi97,oset98}.

One of the main goals of such studies is to explore how the presence
of the new degree of freedom (strangeness) alters and broadens the
knowledge achieved from 
conventional nuclear physics. Several features of the
$\Lambda$ single-particle properties in the nucleus, being 
essentially different from  those of the nucleon, have clearly
emerged from these efforts. It is well accepted nowadays that the
depth of the $\Lambda$-nucleus potential is around $-30$ MeV,
which is 20 MeV less attractive than the
corresponding nucleon-nucleus one. The spin-orbit splittings of single
particle levels in $\Lambda$ hypernuclei were found to be much smaller
than their nucleonic counterparts\cite{bruck78}, typically more
than one order of magnitude . Moreover,
the $\Lambda$, contrary to the nucleon, maintains its single-particle
character even for states well below the Fermi surface \cite{pile91,hase96} indicating
a weaker interaction with other nucleons. Studies of the mesonic weak
decay of light $\Lambda$ hypernuclei \cite{moto91,straub93,kuma95}
have shown that the data
\cite{szyman91} clearly favour $\Lambda$-nucleus potentials which
show a repulsion at short distances. This seems also to be a
characteristic of the $\Sigma$-nucleus potential for light
$\Sigma$-hypernuclei \cite{harada90}, which reproduces the
recently measured bound $\Sigma^+$ state in $^{4}_{\Sigma}$He with the
in-flight $^4$He$(K^-,\pi^-)$ reaction \cite{nagae98}. This experiment 
confirms, with new and better statistics, the earlier results from the
$^4$He(stopped $K^-,\pi^-)$ reaction \cite{hayano89}.

Attempts to derive the hyperon properties in a nucleus have followed
several approaches. A $\Lambda$-nucleus potential of Woods-Saxon type
reproduces reasonably well the measured $\Lambda$ single-particle
energies of medium to heavy hypernuclei
\cite{bouy76,dover80,moto88}.
Non localities and density dependent effects, included in
non-relativistic Hartree-Fock calculations using Skyrme
hyperon-nucleon ($YN$)
interactions \cite{yama88,mille88,fernan89,lansk97}, improve the
overall fit to the single-particle binding energies.  The properties
of hypernuclei have also been studied in a relativistic framework,
such as Dirac phenomenology \cite{brock81,chiap91} or
relativistic mean field theory
\cite{mares89,mares94,lomba95,glende93,inei96,suga94,ma96,yama85}.

Microscopic calculations, which aim at relating the hypernuclear
observables to the bare $YN$ interaction, are also available. This
approach is especially interesting because it can be used to put
further constraints on the $YN$ interactions, which are not completely
determined by the limited amount of scattering data due, essentially,
to the
experimental difficulties associated with the short lifetime of hyperons
and low intensity beam fluxes. Microscopic hypernuclear structure
calculations are performed with an effective $YN$ interaction ($G$-matrix)
obtained from the bare $YN$ potential through a
Bethe-Goldstone equation. The comparison with data may therefore further help in
constraining the $YN$ potentials.
In several microscopic calculations a Gaussian
parametrizations of the $G$-matrix calculated in nuclear matter at an
average density \cite{yama85,yama90,yama92,yama94} was employed. 
Furthermore, a $G$-matrix obtained directly in finite nuclei was used to
study the single-particle energy levels of various hypernuclei
\cite{hald93} or as an effective interaction in a calculation of the 
$^{17}_{\Lambda}$O spectrum \cite{hao93}.

Along similar lines, the authors of Ref.\ \cite{hjort96}
derived microscopically  the $\Lambda$ self--energy in $^{17}_{\Lambda}$O,
starting from realistic hyperon-nucleon interactions. 
The starting point in the latter calculations 
is a nuclear matter $G$-matrix at a fixed
energy and density, which is used to calculate the self--energy for the
finite nucleus including corrections up to second order. 
The obtained self--energy for the $\Lambda$ 
is non-local and depends on the energy of the hyperon. Solving
the Schr\"{o}dinger equation with this self--energy it is possible to 
determine the single-particle energies and wave functions of the bound
hyperon. The approach also provides automatically
the real and imaginary part of the hyperon optical potential
at positive energies and, therefore,
allows to study the hyperon-nucleus scattering properties.
The method was first
employed to study the nucleon and $\Delta$ properties in nuclei
\cite{bbmp92,hmp94} and later applied to calculate the $s$-wave
non-local $\Lambda$ self--energy in $^{17}_{\Lambda}$O, 
from which the single-particle
$\Lambda$ binding energy was obtained \cite{hjort96}.

The aim of the present work is 
to extend the calculations 
of Ref.\ \cite{hjort96} in order to derive 
$s$ and $p$-wave $\Lambda$ single-particle energies for a variety of
$\Lambda$ hypernuclei, from $^{17}_{\Lambda}$O 
to $^{209}_{\Lambda}$Pb. The $\Sigma$ single-particle energies obtained for 
different potentials are also
discussed. The $YN$ potentials employed are the Nijmegen soft-core
\cite{nijmegen} and the J\"ulich \cite{juelich} interactions.
Eventual  differences in single-particle binding energies  
may therefore help in further constraining the $YN$ interactions.

In Sect. \ref{formal}, we present the formalism which consists of
obtaining, first, the nuclear matter $G$-matrix at fixed density and
starting energy parameter in the center of mass frame and, next,
performing a
second order calculation to derive the finite nucleus $G$-matrix from
which one can obtain the hyperon self--energy for the different
single-particle 
orbits.
This self--energy can in turn be used to define a single-particle
potential in
a Schr\"odinger equation in order to obtain the corresponding single-particle
binding energies. Nuclear matter results and a discussion on the
convergence of our
method is presented in Sect. \ref{results1}. An important point raised in
the latter section is that the single-particle hyperon energies calculated
at first order with a nuclear matter $G$-matrix depend quite strongly on 
the fixed starting energy and density employed. The $\Lambda$
single-particle 
energies for a variety of
hypernuclei are shown in Sect. \ref{results2}, where local
equivalent Woods-Saxon potentials to represent our non-local
self--energy are also derived. Finally, our conclusions are presented in
Sect. \ref{conclusions}.

\section{Formalism}
\label{formal}
In this section we present the formalism to obtain the hyperon single-particle
energies in finite hypernuclei using an effective interaction
derived microscopically from realistic $YN$ interactions, for which we 
take the Nijmegen soft core \cite{nijmegen} and J\"ulich\cite{juelich} 
potentials. 
Although the formalism was already described in Ref. \cite{hjort96}, 
part of it will be repeated here in order to set up the notation used 
later in the description of the results.
In the first place, we solve the $G$-matrix in nuclear matter at a
fixed density, center of mass momentum and
energy. Next, we transform this $G$-matrix into an effective
interaction for the finite hypernucleus from which we can obtain the
hyperon self--energy. Finally, we use this self--energy in a Schr\"odinger
equation to derive the single-particle energies and the corresponding
wave functions of the bound states. 

\subsection{Nuclear matter $YN$ $G$-matrix}

The nuclear matter $YN$ $G$-matrix is solved in momentum space and the 
two-particle $YN$ states are defined in terms of 
relative and the center-of-mass
momenta, {\bf k} and {\bf K}, given by

\begin{eqnarray}
 {\bf k} &=&\frac{M_N{\bf k}_Y-M_Y{\bf k}_N}{M_N+M_Y}, \nonumber \\   
  {\bf K} &=&{\bf k}_N+{\bf k}_Y \ , \nonumber   
\end{eqnarray}
where ${\bf k}_N$ and ${\bf k}_Y$ are the nucleon and hyperon momenta,
respectively.
Using an angle-averaged Pauli operator, we perform a partial wave
decomposition of the Bethe-Goldstone equation, which, in terms of the 
quantum numbers of the relative and center-of-mass motion (RCM), is 
written as
\begin{eqnarray}
\lefteqn{\left\langle (Y''N)k''l''KL({\cal J})S''T_z\right |
      G(\omega_{NM})\left | (YN)klKL({\cal J})ST_z \right\rangle
      =} \hspace{2cm}\nonumber\\
  && \left\langle (Y''N)k''l''KL({\cal J})S'' T_z\right |
      V\left | (YN)klKL({\cal J})ST_z \right\rangle
      \nonumber \\
      &\quad +& {\displaystyle
      \sum_{l'}\sum_{S'}\sum_{Y'=\Lambda\Sigma}\int k'^{2}dk'}
      \left\langle (Y''N)k''l''KL({\cal J})S''T_z\right |
      V\left | (Y'N)k'l'KL({\cal J})S'T_z \right\rangle \nonumber \\
      && \times \frac{Q(k',K)}{\omega_{NM} -\frac{K^2}{2(M_N+M_{Y^{'}})} -
      \frac{k'^2(M_N+M_{Y^{'}})}{2M_NM_{Y^{'}}}-M_{Y^{'}}+M_Y}
\nonumber \\
      && \times\left\langle (Y'N)k'l'KL({\cal J})S'T_z\right |
      G(\omega_{NM})\left | (YN)klKL({\cal J})ST_z
\right\rangle \ ,
   \label{eq:gmat}
\end{eqnarray}
where $Q$ is the nuclear matter Pauli operator, $V$ is the $YN$ potential
and $\omega_{NM}$
is the nuclear matter starting energy which corresponds to the sum of 
non-relativistic single-particle energies of the interacting nucleon and 
hyperon. 
Note that kinetic energies are used in the intermediate $Y^\prime N$ states
and $M_{Y}-M_{Y^\prime}$ accounts for the mass difference of the initial 
and intermediate hyperon.
The
variables $k$, $k'$, $k''$ and $l$, $l'$, $l''$ denote relative momenta
and angular momenta, respectively, while $K$ and $L$ 
are the quantum numbers of the center-of-mass motion. 
The total angular momentum,  spin and isospin
projection of the $YN$ pair are denoted by ${\cal J}$, $S$ and
$T_z$, respectively.

\subsection{The hyperon single-particle potential $U_{Y}$ in nuclear matter}

In the Brueckner-Hartree-Fock approach 
the hyperon single-particle potential $U_{Y}$ is obtained
self-consistently by the following sum of
diagonal $G$-matrix elements:
\begin{equation}
     U_{Y}(k_{Y})=
   \int_{k_{N}\leq k_{F}}d^3k_{N}\left\langle Y{\bf k}_{Y},N{\bf k}_{N}\right
| G(\varepsilon_N(k_N)+\varepsilon_{Y}(k_Y))\left | Y{\bf
k}_{Y},N{\bf k}_{N}\right\rangle \ ,  
\end{equation} 
where $\varepsilon_{N(Y)} (k_{N(Y)}) = k_{N(Y)}^2/(2 M_{N(Y)}) +
U_{N(Y)}(k_{N(Y)})$ is
the single-particle energy of the nucleon (hyperon). Using the
partial wave decomposition of the $G$-matrix, the single-particle 
potential $U_Y$  can be rewritten as
\begin{eqnarray}
   U_{Y}(k_{Y})&=&
\frac{(1+\xi_{Y})^3}{2(2t_{Y}+1)}\sum_{{\cal J},l,S,T}(2{\cal J}+1)(2T+1) 
\nonumber \\
& & \times \int_{0}^{k_{max}}k^2dk f(k,k_{Y})
   \left\langle Y N; k lST_z\right | G^{\cal J}(\varepsilon_N(k_N)+\varepsilon_{Y}(k_Y))\left | 
Y N; klST_z\right\rangle  \ ,
\label{eq:uy}
\end{eqnarray}
where an average over the hyperon spin and isospin ($t_Y$) has been performed and the weak center-of-mass
dependence of the $G$-matrix has been neglected.
In Eq. (\ref{eq:uy}), $k$ is the relative momentum of the $YN$ pair, 
$\xi_Y=M_N/M_Y$, $k_{max}$ is given by
\begin{equation}
   k_{max} = \frac{k_{F}+\xi_{Y}k_{Y}}{1+\xi_{Y}} \nonumber  
\end{equation}
and the weight function $f(k,k_Y)$ by 
\begin{equation}
f(k,k_{Y})= \left\{ \begin{array}{cl} 1 & \mbox{for $ k\leq
\frac{k_{F}-\xi_{F}k_{Y}}{1+\xi_{Y}} $ }, \\ 0 & \mbox{for $
|\xi_{Y}k_{Y}-(1+\xi_{Y})k| > k_{F} $ }, \\
\frac{k_{F}^2-[\xi_{Y}k_{Y}-(1+\xi_{Y})k]^2}{4\xi_{Y}(1+\xi_{Y})k_{Y}k
} & \mbox{otherwise} 

\end{array} \right. \nonumber 
\end{equation} 

In the discussion, we will associate the binding energy of the
hyperon to its single-particle energy:
\begin{equation}
B_Y(k_Y)\equiv \varepsilon(k_Y)= \frac{k_Y^2}{2 M_Y} + U_Y(k_Y)
\label{eq:binding}
\end{equation}


\subsection{Evaluation of the hyperon self--energy}

\begin{figure}
       \setlength{\unitlength}{1mm}
       \begin{picture}(100,180)
       \put(25,-50){\epsfxsize=12cm \epsfbox{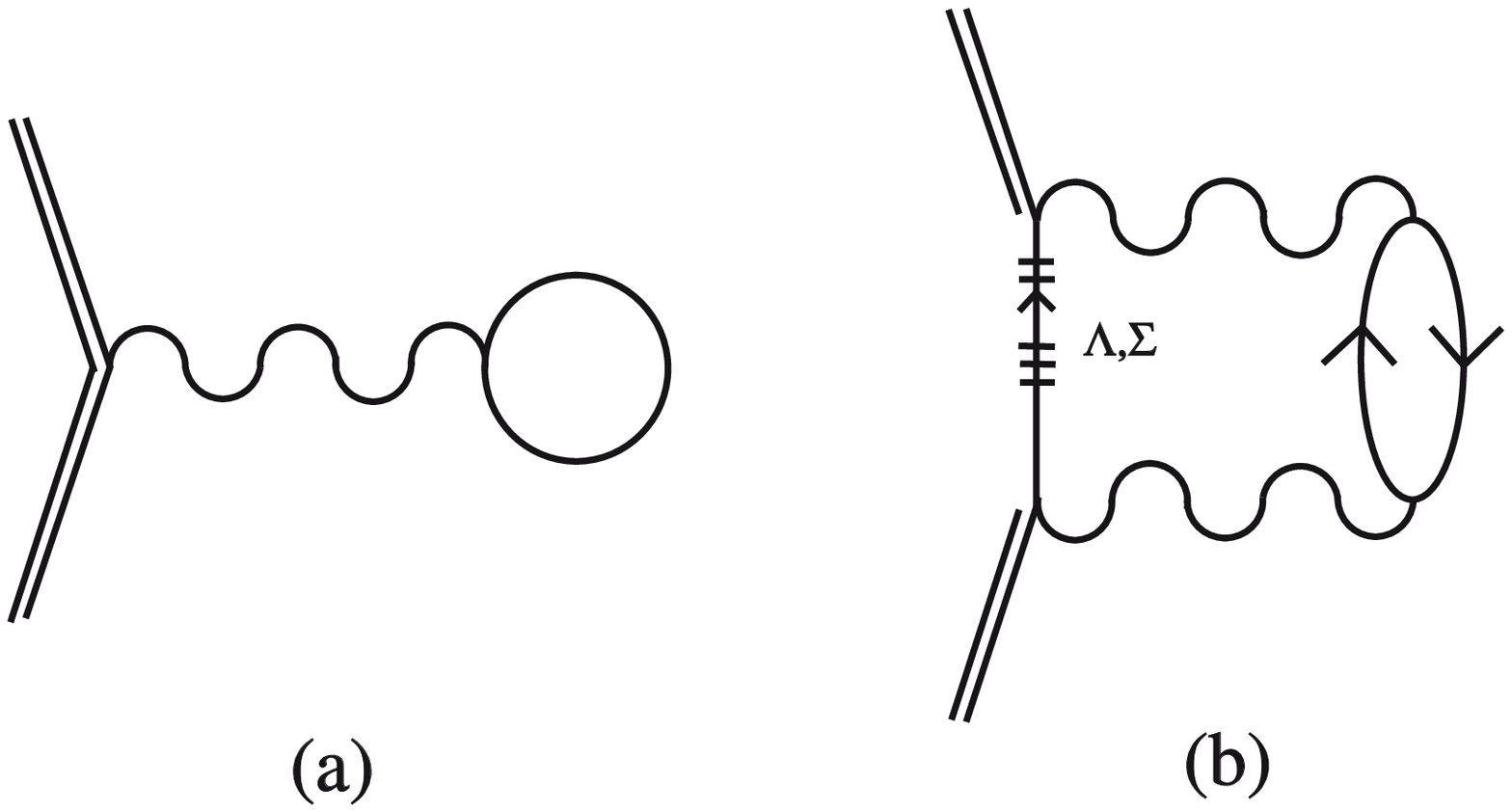}}
       \end{picture}
   \caption{Diagrams through second order in the interaction $YN$ $G$
   (wavy line) included in the evaluation of the hyperon self--energy.
Diagram
   (a) is the first order term, while
   (b) is the second order $2p1h$ correction.}
   \label{fig:fig1}
\end{figure}

The self--energy of the $\Lambda$ or $\Sigma$ hyperon in a finite
hypernucleus can be obtained, in the Hartree-Fock scheme, using a
finite nucleus $G$-matrix as an effective $YN$ interaction. 
However, our $G$-matrix
has been obtained in nuclear matter and, in particular, the
intermediate propagator (that involves the Pauli operator and energy
denominator) is very different than the corresponding one in a finite
hypernucleus.  One can, nevertheless, find the appropriate finite 
nucleus $G$-matrix, $G_{FN}$, by relating it to the 
nuclear matter $G$-matrix through the following integral equation written 
in schematic form:
\begin{eqnarray}
G_{FN} &= &G + G \left[ \left( \frac{Q}{e} \right)_{FN} 
-\left( \frac{Q}{e} \right)_{NM} \right] G_{FN} \nonumber \\
&= &G + G \left[ \left( \frac{Q}{e} \right)_{FN} 
-\left( \frac{Q}{e} \right)_{NM} \right] G  \nonumber \\
&& + G \left[ \left( \frac{Q}{e} \right)_{FN}  
-\left( \frac{Q}{e} \right)_{NM} \right] 
 G \left[ \left( \frac{Q}{e} \right)_{FN} 
-\left( \frac{Q}{e} \right)_{NM} \right] G + \dots  \ ,
\label{eq:gexpan}
\end{eqnarray}
which involves the nuclear matter $G$-matrix (labelled $G$ throughout the
text) and the difference between
the finite nucleus and nuclear matter propagators. The latter account
for the relevant intermediate states.
In the practical calculations we will approximate
the expansion up to the second order in the nuclear matter $G$-matrix
\begin{equation}
G_{FN} \simeq G + G \left[ \left( \frac{Q}{e} \right)_{FN} 
-\left( \frac{Q}{e} \right)_{NM} \right] G \ . 
\label{eq:gexpan2}
\end{equation}
Therefore, in the evaluation of the hyperon
self--energy we take into account 
the diagrams
displayed in Fig.~\ref{fig:fig1}, where the wiggly interaction lines 
represent the nuclear matter $G$-matrix.  Diagram \ref{fig:fig1}(a)
represents the first term on the right hand side
of Eq. (\ref{eq:gexpan2}) which, by analogy to the nuclear case, will be
refered to as the Hartree-Fock contribution. Diagram \ref{fig:fig1}(b)
stands for the
second order correction, where the
intermediate propagator has to be viewed as the difference of
propagators appearing in Eq. (\ref{eq:gexpan2}).

We will consider the incoming (outcoming) hyperon as a plane wave and
the nucleon hole states as harmonic oscillator ones, so the two-body 
wave function will be a mixed representation
of both single-particle states given by
\begin{equation}
\hspace{-0.3cm}\left | (n_hl_hj_ht_{z_h})(k_Yl_Yj_Yt_{z_Y})JT_z \right \rangle=
    \int k_h^2dk_hR_{n_hl_h}(\alpha k_h)
    \left | (k_hl_hj_ht_{z_h})(k_Yl_Yj_Yt_{z_Y})JT_z \right \rangle,
\label{eq:trans1}
\end{equation}
where $n_hl_hj_ht_{z_h}$ and $k_Yl_Yj_Yt_{z_Y}$ are the quantum numbers of the nucleon hole state
and the hyperon state, respectively. Further, $\alpha$ is the oscillator parameter appropiate 
to describe the single-particle wave functions of the bound
nucleons in the nuclear core. It is defined as
\begin{equation}
   \alpha=\frac{\hbar c}{\sqrt{M_{N}\hbar \omega}}
\label{eq:alfa}
\end{equation} 
with $\hbar \omega$ chosen as the following function of the mass number
\begin{equation} 
   \hbar \omega=45A^{-\frac{1}{3}}-25A^{-\frac{2}{3}} \ .
\label{eq:hachbar}
\end{equation} 

Typical matrix elements needed in the calculation are
\begin{equation}
   \left\langle (k_Yl_Yj_Yt_{z_Y})(n_hl_hj_ht_{z_h})JT_z\right |
    G | (k_{Y}'l_{Y}j_{Y}t_{z_{Y}})(n_{h}l_{h}j_{h}
t_{z_{h}})JT_z \rangle \ ,
\label{eq:g1}
\end{equation}
for the Hartree-Fock diagram of Fig.~\ref{fig:fig1}(a), or
\begin{equation}
   \left\langle (k_Yl_Yj_Yt_{z_Y})(n_hl_hj_ht_{z_h})JT_z \right |
    G\left | (YN)klKL({\cal J})ST_z \right\rangle  ,
\label{eq:g2}
\end{equation}
appearing in the second-order diagram of Fig.~\ref{fig:fig1}(b).

Using Eq. (\ref{eq:trans1}), the mixed representation states appearing
in  Eqs. (\ref{eq:g1}) and 
(\ref{eq:g2}) can be expressed in terms of momentum and angular
momentum variables in the laboratory frame. 
Then, with appropriate transformation coefficients  \cite{kkr79,wc72}, 
one can express
the two-body states with laboratory coordinates in terms of the
variables in the RCM system used in the solution of the 
$G$-matrix\footnote{Note the distinction between
$k_a$ and $k$ and $l_a$ and $l$. With the notation $k_a$
or $l_a$ we will refer to the quantum numbers of the single-particle
state, whereas $l$ or $k$ without subscripts refer to the coordinates  
of the relative motion.}
 
\begin{equation}
   \begin{array}{ll}
     &\\
     \left | (k_al_aj_at_{z_a})(k_bl_bj_bt_{z_b})JT_z\right \rangle =&
      {\displaystyle \sum_{lL\lambda S{\cal J}}}\int k^{2}dk\int K^{2}dK
      \left\{\begin{array}{ccc}
      l_a&l_b&\lambda\\\frac{1}{2}&\frac{1}{2}&S\\
      j_a&j_b&J\end{array}
      \right\}\\&\\
      &\times (-1)^{\lambda +{\cal J}-L-S}
      \hat{{\cal J}}\hat{\lambda}^{2}
      \hat{j_{a}}\hat{j_{b}}\hat{S}
      \left\{\begin{array}{ccc}L&l&\lambda\\S&J&{\cal J}
      \end{array}\right\}\\&\\
      &\times \left\langle klKL| k_al_ak_bl_b\right\rangle
      \left | klKL({\cal J})SJT_z\right \rangle ,
   \end{array}
\end{equation}
where $\hat{x}=\sqrt{2x+1}$ and  $\left\langle klKL|
k_al_ak_bl_b\right\rangle$ are
the transformation coefficients from the RCM system to the laboratory 
system.

We can construct now the expressions for the
various diagrams considered in this work. The first order term 
of Fig. \ref{fig:fig1}(a) yields
a real and energy-independent contribution to the self--energy given by
\begin{eqnarray}
\lefteqn{  {\cal V}_{HF}(k_{Y}k_{Y}'l_{Y}j_{Y}
   t_{z_{Y}})\quad = \quad \frac{1}{\hat{j_{Y}}^2}
  {\displaystyle \sum_{J}\sum_{n_hl_hj_ht_{z_h}}}\hat{J}^2}\hspace{1.5cm}
\nonumber \\
  & \times & \left\langle
(k_{Y}l_{Y}j_{Y}t_{z_{Y}})
  (n_hl_hj_ht_{z_h})JT_z \right |
   G\left | (k_{Y}l_{Y}j_{Y}t_{z_{Y}})
   (n_hl_hj_ht_{z_h})JT_z \right \rangle ,
\label{eq:hf}
\end{eqnarray} 
where $k_{Y}$($k_{Y}'$)$l_{Y}$$j_{Y}$$t_{z_{Y}}$ are
the quantum numbers of the incoming(outcoming) hyperon. \\
The computation of the contribution coming from the 
two-particle-one-hole ($2p1h$) diagram of Fig.~\ref{fig:fig1}(b)
requires a little more work. First, we evaluate the imaginary part of the 
second term in Eq.\ (\ref{eq:gexpan2}). This term has an
an explicit energy dependence. It reads   
\begin{eqnarray}
\lefteqn{{\cal W}^{(1)}_{2p1h}(k_{Y}k_{Y}'l_{Y}j_{Y}
      t_{z_{Y}}\omega) \ = 
      {\displaystyle -\frac{1}
      {\hat{j_{Y}}^2}\sum_{n_hl_hj_ht_{z_h}}
      \sum_{J}\sum_{lLS{\cal J}}\sum_{Y'=\Lambda\Sigma}\int k^{2}dk
      \int K^{2}dK\hat{J}\hat{T}}}\hspace{1cm}\nonumber \\
      &\times& \left\langle (k_{Y}'l_{Y}j_{Y}
      t_{z_{Y}})(n_hl_hj_ht_{z_h})JT_z\right |
      G\left | (Y'N)klKL({\cal J})SJT_z \right \rangle\nonumber \\
      &\times& \left\langle (Y'N)klKL({\cal J})SJT_z \right | G
      \left | (k_{Y}l_{Y}j_{Y}
      t_{z_{Y}})(n_hl_hj_ht_{z_h})JT_z \right \rangle \nonumber
      \\ &\times& \pi\delta
      \left(\omega +\varepsilon_h -\frac{K^2}{2(M_N+M_{Y^{'}})} -
      \frac{k^2(M_N+M_{Y^{'}})}{2M_NM_{Y^{'}}}-M_{Y^{'}}+M_Y \right) \ ,
   \label{eq:2p1h}
\end{eqnarray} \\
where $\omega$ is the energy of the hyperon measured with respect
to the hyperon rest mass. The single-hole energies $\varepsilon_{h}$ have
been taken equal to the
experimental single-particle energies in most of the nuclei studied (e.g.,
$^{12}$C, $^{16}$O, $^{40}$Ca) and have been calculated from a
Woods-Saxon potential with Spin-Orbit and Coulomb terms appropriately fitted
in other cases (e.g., $^{90}$Zr and $^{208}$Pb). The
quantities
$klKL({\cal J})SJT_z$ are the
quantum numbers of the intermediate $Y'N$ states. Next,
we obtain the real part through a 
dispersion relation
\begin{equation}
   {\cal V}^{(1)}_{2p1h}(k_{Y}k_{Y}'l_{Y}j_{Y}
   t_{z_{Y}} \omega)=
   \frac{P}{\pi} \int_{-\infty}^{\infty}
   \frac{{\cal W}_{2p1h}(k_{Y}k_{Y}'l_{Y}
    j_{Y}
    t_{z_{Y}} \omega')}{\omega'-\omega} d\omega',
    \label{eq:disprel}
\end{equation}
where $P$ means a principal value integral. 

Finally, we must subtract the $2p1h$  correction term 
coming from the nuclear matter intermediate propagator
(third term in Eq.\ (\ref{eq:gexpan2}) ). It reads
\begin{eqnarray}
\lefteqn{\hspace{-0.5cm}{\cal V}^{(2)}_{2p1h}(k_{Y}k_{Y}'l_{Y}   
      j_{Y}t_{z_{Y}}) \ = \
      {\displaystyle \frac{1}
      {\hat{j_{Y}}^2}\sum_{n_hl_hj_ht_{z_h}}
      \sum_{J}\sum_{lLS{\cal J}}\sum_{Y'=\Lambda\Sigma}\int k^{2}dk
      \int K^{2}dK\hat{J}\hat{T}}}\hspace{0.5cm}\nonumber \\
      &\times& \left\langle (k_{Y}'l_{Y}j_{Y} 
      t_{z_{Y}})(n_hl_hj_ht_{z_h})JT_z\right | 
      G\left | (Y'N)klKL({\cal J})SJT_z \right \rangle\nonumber \\
      &\times& \left\langle (Y'N)klKL({\cal J})SJT_z \right | G
      \left | (k_{Y}l_{Y}j_{Y}
      t_{z_{Y}})(n_hl_hj_ht_{z_h})JT_z \right \rangle \nonumber
      \\
      &\times& Q(k,K)\left(\omega_{NM} -\frac{K^2}{2(M_N+M_{Y^{'}})} -
      \frac{k^2(M_N+M_{Y^{'}})}{2M_NM_{Y^{'}}}-M_{Y^{'}}+M_Y\right)^{-1} \
,
   \label{eq:vc}
\end{eqnarray}
where $Q$ is the nuclear matter Pauli operator
and $\omega_{NM}$ is the nuclear matter starting
energy.
This term only contributes to the real part
of the hyperon self--energy and avoids the double counting over
intermediate $Y^\prime N$ states contained already in the nuclear
matter $G$-matrix of the Hartree-Fock contribution ${\cal V}_{HF}$.

In summary, the self--energy of the hyperon reads
\begin{equation}
    \Sigma(k_{Y}k_{Y}'l_{Y}j_{Y}\omega)=
    V(k_{Y}k_{Y}'l_{Y}j_{Y}\omega)+
    iW(k_{Y}k_{Y}'l_{Y}j_{Y}\omega),
    \label{eq:self_ener}
\end{equation}
with the real part given by
\begin{equation}
    V(k_{Y}k_{Y}'l_{Y}j_{Y}\omega)=
       {\cal V}_{HF}(k_{Y}k_{Y}'l_{Y}j_{Y})+
       {\cal V}^{(1)}_{2p1h}(k_{Y}k_{Y}'l_{Y}j_{Y}
        \omega)-
       {\cal V}^{(2)}_{2p1h}(k_{Y}k_{Y}'l_{Y}j_{Y}
       )
       \label{eq:realV}
\end{equation}
and the imaginary part by

\begin{equation}
    W(k_{Y}k_{Y}'l_{Y}j_{Y}\omega)=
       {\cal W}^{(1)}_{2p1h}(k_{Y}k_{Y}'l_{Y}j_{Y}\omega).
       \label{eq:imV}
\end{equation}


The self--energy can then be inserted as a single-particle
potential in a Schr\"o\-din\-ger equation in order to investigate
bound and scattering states of a  hyperon
in a finite nucleus. The different approximations to the
self--energy, i.e., whether we include the $2p1h$
contribution
or not, result in different single-particle hamiltonians. We solve
the Schr\"o\-din\-ger equation by diagonalizing the corresponding
single-particle hamiltonian in a complete basis within a spherical box
of radius $R_{box}$ following the procedure outlined in \cite{hjort96}. 
This method  is especially suitable fon non-local potentials defined either in coordinate or 
in momentum space \cite{bbmp92,hmp94}.

\section{Results and Discussion}
\label{results}
In this section we present and discuss results for $\Lambda$
and $\Sigma$ hypernuclei
using two realistic interactions: 
Nijmegen soft core \cite{nijmegen} and J\"ulich B \cite{juelich}.

\subsection{Dependence of results on the starting energy}
\label{results1}

As described before, our method provides the binding energies of the
different hyperon orbits in finite hypernuclei starting from a
$G$-matrix calculated in nuclear matter in the $YN$
center-of-mass frame at
fixed starting energy $\omega_{NM}$ and Fermi momentum $k_F$.
By adding the $2p1h$ correction to the Hartree-Fock term one incorporates,
up to second order in the nuclear matter $G$-matrix, the correct
energy dependence and Pauli blocking factor in the finite nucleus.
Therefore, the complete calculation ($HF+2p1h$) has to be
viewed as a Hartree-Fock approach which uses a 
effective interaction derived microscopically with the appropriate density and energy
dependence of the hypernucleus under study.
This is in contrast to previous calculations 
\cite{yama85,yama90,yama92,yama94} where the determination of 
the finite hypernucleus effective interaction from
the nuclear matter $G$-matrix implied a sort of
average over the nuclear density. In these works several local and
energy independent effective $YN$ interactions of Gaussian form ($YNG$)
were derived by parametrizing the corresponding nuclear matter
$G$-matrices obtained from various $YN$ potentials.
The parametrization of the $G$-matrix into a local effective
interaction $YNG$ to be used in finite hypernuclei calculations
required the use of an appropiate value of the Fermi momentum $k_F$.
This value was determined, for each
nucleus, by averaging the corresponding nuclear density weighted by
the modulus squared of the $\Lambda$
single-particle wave function of the single-particle
level under study. The parameters of the effective $YNG$ interaction 
were adjusted to
reproduce the
$\Lambda$ potential energy $U_\Lambda(0)$ in nuclear matter at the
average value of $k_F$.
With these parametrizations, $\Lambda$
single-particle energies and excited hypernuclear levels in several
$\Lambda$
hypernuclei were obtained through a shell-model calculation, with the
aim of learning about the bare $YN$ interaction.

It therefore seems appropriate to explore, using our method, how much
the hyperon single-particle energy depends on the starting energy (and
density) 
of the nuclear matter
$G$-matrix used in the calculation. This will allow us to assess
how reliable energy independent effective interactions obtained
from local density averages might be.

\begin{figure}
       \setlength{\unitlength}{1mm}
       \begin{picture}(100,180)
       \put(5,10){\epsfxsize=14cm \epsfbox{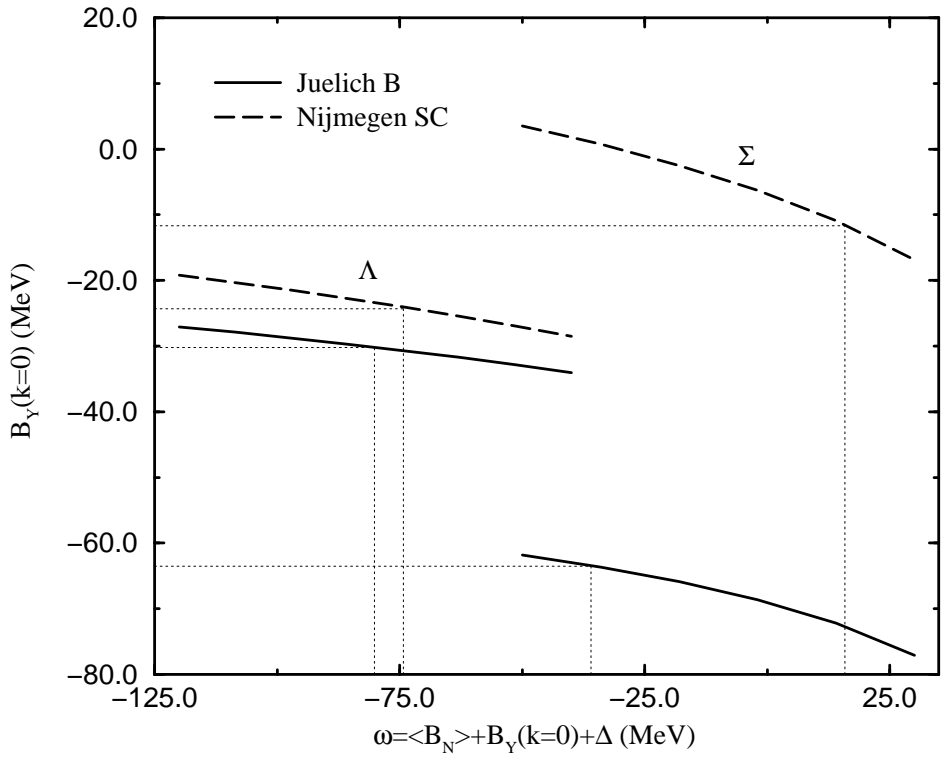}}
       \end{picture}
   \caption{Dependence of the hyperon binding energy $B_{Y}(k=0)$ in nuclear matter 
            on the starting 
            energy $\omega$. The curves on the left are for the $\Lambda$,
            whereas the ones on the right are for the $\Sigma$.
Long-dashed (full)
            lines correspond to Nijmegen soft core (J\"ulich B)
interaction. The dotted lines show the position of the self-consistent
solution for $B_Y(k=0)$. }
    \label{fig:binener}
\end{figure}

Let us first show, in Fig. \ref{fig:binener}, the binding energy
$B_Y(k=0)$ of
a $\Lambda$ (curves on the left) or a $\Sigma$ (curves on the right)
in nuclear matter at $k_F=1.36$ fm$^{-1}$ as a function of the starting
energy parameter $\omega=\omega_{NM}+\Delta = <B_N> + B_Y(k=0) +
\Delta$, 
where $<B_N>=-50$
MeV is an average of
the nucleon binding energy over the Fermi sea and $\Delta=M_Y-M_\Lambda$.
The long-dashed (full) lines are for the Nijmegen soft core (J\"ulich B)
interaction. An estimate of the
self-consistent solution is obtained where the line 
$\omega=\omega_{NM}+\Delta = <B_N> + B_Y(k=0) +
\Delta$ crosses the calculated values of $B_Y(k=0)$. This is indicated by
the dotted lines in the figure.
In the case of the Nijmegen interaction we obtain  
$\omega=-74.3$ MeV ($B_\Lambda(0)=-24.3$ MeV) for the
$\Lambda$ and $\omega=15.8$ MeV ($B_\Sigma(0)=-11.7$ MeV) for the
$\Sigma$, 
whereas, in the case
of the J\"ulich interaction,  $\omega=-80.2$ MeV for the
$\Lambda$ ($B_\Lambda(0)=-30.2$ MeV) and $\omega=-36.0$ MeV 
($B_\Sigma(0)=-63.5$ MeV) for the $\Sigma$.

Several features emerge from Fig. \ref{fig:binener}. First,
the $\Sigma$ hyperon is unrealistically overbound in nuclear matter 
by the J\"ulich
interaction. It is therefore necessary to readjust the parameters
of this interaction if
one wants to use it in shell model calculations of $\Sigma$
hypernuclei.
Secondly, we observe that the energy dependence of $U_Y(k=0)$ is
slightly stronger in the case of the Nijmegen interaction, especially
for the $\Sigma$ hyperon which is more sensitive to the
$\Sigma N - \Lambda N$ coupling
because the starting energy is closer to the energies of the
intermediate $\Lambda N$ states (which propagate with the kinetic energy
spectrum). Finally, we observe that the $\Lambda$ binding energy 
varies at most by 10 MeV in a starting energy range of 80 MeV,
while the variation of the $\Sigma$ binding
amounts to twice as much. As we will see below, this has consequences
in the results for finite hypernuclei.

In Tables \ref{tab:olambda} and \ref{tab:osigma} we show the binding
energy of the $\Lambda$ and $\Sigma^0$, respectively, in $^{17}_Y$O.
The
columns denoted by $HF$ correspond to our lowest order calculation
(see
Eq.~(\ref{eq:hf})) which uses, as effective interaction, the nuclear
matter $G$-matrix calculated in the $YN$ center of mass frame
at fixed energy (shown in the first column) and density ($k_F=1.36$
fm$^{-1}$).
Columns labelled $(HF + 2p1h)$ include the $2p1h$ corrections (see
Eqs.~(\ref{eq:disprel}) and (\ref{eq:vc})) to bring the nuclear
matter $G$-matrix to the finite nucleus one, with the proper
energy
and density dependence. We see that the lowest order results depend
quite strongly on the starting energy used, especially in the case
of the $\Sigma$ hyperon as would be expected from the nuclear matter
results shown in Fig. \ref{fig:binener}. However, it is worth noticing
how,
no matter what
starting energy is used in solving the nuclear matter $G$-matrix, the
corrected
calculation ($HF+2p1h$) ends up giving practically the same result for the
hyperon
binding
energy. Particularly stable are the results for the $\Lambda$ hyperon.
This weaker energy dependence, seen already in the nuclear matter
results of Fig. \ref{fig:binener}, is due to the fact that the energies
involved in the calculation lie further away from the
intermediate $YN$ states, which propagate with kinetic energy,
and therefore the strong $\Lambda N-\Sigma N$ coupling is less
pronounced.

In Tables \ref{tab:dlambda} and \ref{tab:dsigma} we show the binding
energy of the $\Lambda$ and $\Sigma^0$, respectively, in $^{17}_Y$O
using nuclear matter $G$-matrices calculated at several values of the Fermi momentum
and a
fixed value of the starting energy ($\omega=-50$ for the $\Lambda$ and
$\omega=0$ for the $\Sigma$).
The lowest order calculation for the hyperon single-particle energy, shown
in the second column, depends quite strongly on the value of $k_F$.
However, one finds again that, when the $2p1h$ correction is included
to incorporate the proper intermediate propagator of the finite
nucleus, the results nicely converge to practically the same value, no
matter what was the density used in the solution of the nuclear matter
$G$-matrix.

These results  are interesting because they confirm that
the finite nucleus $G_{FN}$-matrix is already well 
approximated by the second order term 
in the expansion in terms of the nuclear matter $G$-matrix, see
the second and third terms in Eq.\ (\ref{eq:gexpan2}).
The correction, whose size depends on the starting energy or Fermi
momentum used in the solution of the nuclear matter $G$-matrix, already
leads to practically the same value for the hyperon single-particle energy.
Higher order terms could only help in bringing the results closer than
what they already are.
Moreover, our results also show that in some cases the correction is
quite
appreciable, not only for the $\Sigma^0$ binding energies shown in
Tables \ref{tab:osigma} and \ref{tab:dsigma}, but also for the
$\Lambda$ energies in
the case of the Nijmegen interaction. Therefore, if the correction is
taken only approximately through an averaged nuclear matter $G$-matrix 
\cite{yama85,yama90,yama92,yama94} it may not lead to the proper
effective interaction in the finite nucleus one is studying. These
words of caution are particularly relevant in the case of the $\Sigma$
hyperon where the corrections are very large.

We note that the $\Lambda$ single-particle energy obtained in the case of the
Nijmegen soft core interaction is in excellent agreement with that
obtained by Halderson (see column 3 in Fig. 7 of Ref.
\cite{hald93}), where the $G$-matrix
was calculated directly in the finite nucleus for various Nijmegen
interactions. 
Our method must be viewed as an alternative
way of building up a finite nucleus
effective interaction. 
It was already
shown there that the Pauli corrections, which are a source of nucleus
dependence, were very large for the Nijmegen soft core potential. This
again supports our believe that calculations based on nuclear matter
$G$-matrices at an average density will carry uncertainties tied to
the chosen value of the Fermi momentum. In particular, the whole
purpose of using these microscopic calculations to constrain the
$YN$ force cannot be achieved if a different value of
$k_F$ is used for each $YN$ interaction when studying the same 
hypernucleus \cite{yama92}.

In the studies of single-particle binding energies below,
we will refrain from a study of the $\Sigma$ binding energy
since the results for $^{17}_{\Sigma}$O give single-particle binding 
energies which are much too
attractive. Several analysis of $\Sigma^{-}$ atomic data \cite{sig1,sig2}
suggest a $\Sigma$ well depth similar to that of the $\Lambda$
\cite{batty79,batty81} and more recent analysis \cite{batty94,mares95} did
not discard $\Sigma$-nucleus potentials showing an inner repulsion.
Moreover, $\Sigma$ hypernuclear spectra from $(K^-,\pi^-)$ reactions
suggest a relatively shallow $\Sigma$-nucleus potential \cite{dover89}.
Therefore, although the amount of data is limited, there is no 
experimental evidence
for such strongly bound $\Sigma$ hyperons. 
This clearly points to a weakness in the present $YN$ interactions,
hinting possibly at a too strong $\Sigma N - \Lambda N$ coupling
in the interactions.

\subsection{$\Lambda$ Single-particle states}
\label{results2}

Once the method is well established and tested for the specific case
of $_{\Lambda}^{17}$O, it is the right moment to study the systematics of the 
$\Lambda$ binding energy through the periodic table. 
To this end, values of the $\Lambda$ single-particle binding energies
 obtained in what has
been called $HF$ and $HF+2p1h$ approximations are reported in
Table \ref{tab:ener} together with the available experimental data.
These binding energies have been
calculated using the energy-independent version with parameter set B
of the J\"ulich potential \cite{juelich}. The results for the Nijmegen
soft core potential \cite{nijmegen} have
not been considered in this section because the corresponding prediction
for the
$\Lambda$ binding energy in nuclear matter is $-23.4$ MeV,
about $7$ MeV weaker than the prediction of the J\"ulich model
($-30.2$ MeV) which agrees well with the extrapolated experimental values
 of the
single-particle $\Lambda$ binding energies with increasing mass number.

The agreement with the experimental data is rather good.  
For convenience in the technicalities of the algorithm we have always
considered 
hypernuclei with a number of nucleons closing a subshell plus a $\Lambda$. 
Unfortunately, experimental data for those nuclei do not always exist and, 
as indicated in the Table \ref{tab:ener},
we have
taken the closest representative nucleus for which the experimental information is
available. Nevertheless, the differences between the calculated and the
experimental values should 
not be associated to this fact but to the approximations used in the calculation
or to the potential itself. 

 For the density and starting energy used to
calculate the $\Lambda$-nucleon $G$-matrix in nuclear matter, which
has been used as effective interaction in our finite nucleus
calculation, it turns out that the $2p1h$ correction is always
attractive. As
discussed in the previous section, if we had used other starting
values for the density or the starting energy  we would have ended up
with different $2p1h$ corrections but with the same value for
$HF+2p1h$.

In agreement with the experimental information, the difference between
 the $p_{3/2}$ and $p_{1/2}$ $\Lambda$ single-particle
 binding energies associated to these partial waves is very
small. Note that the $p_{1/2}$ energy is lower than the $p_{3/2}$. This is
a characteristic of the J\"ulich interaction which yeilds too much attraction
in the $^{3}S_{1}$ partial wave, 
as noted already in \cite{yama92,yama94} where the
$0^{+}$ and $1^{+}$ states of $_{\Lambda}^{4}$He were calculated and
showed to appear in reverse order from the experimental values.

The calculated $\Lambda$ single-particle energies for 
$_{\Lambda}^{209}$Pb appear clearly overbound with respect to the experimental data. 
This is due to the fact that the distortion of
the plane wave associated with the nucleon in the intermediate state of
the $2p1h$ diagram of Fig. \ref{fig:fig1}(b), necessary to ensure its
orthogonalization
to the nucleon hole states, has been taken only approximately. 
The orthogonalization procedure is described in Ref.\ \cite{bbmp92} and
has been optimized for 
the case of $_{\Lambda}^{17}$O. Actually, this
feature is already sizable for $_{\Lambda}^{91}$Zr and in the case of
 $_{\Lambda}^{209}$Pb
leads to a result which is more bound than a $\Lambda$ in nuclear matter.

\begin{figure}
       \setlength{\unitlength}{1mm}
       \begin{picture}(100,180)
       \put(5,10){\epsfxsize=14cm \epsfbox{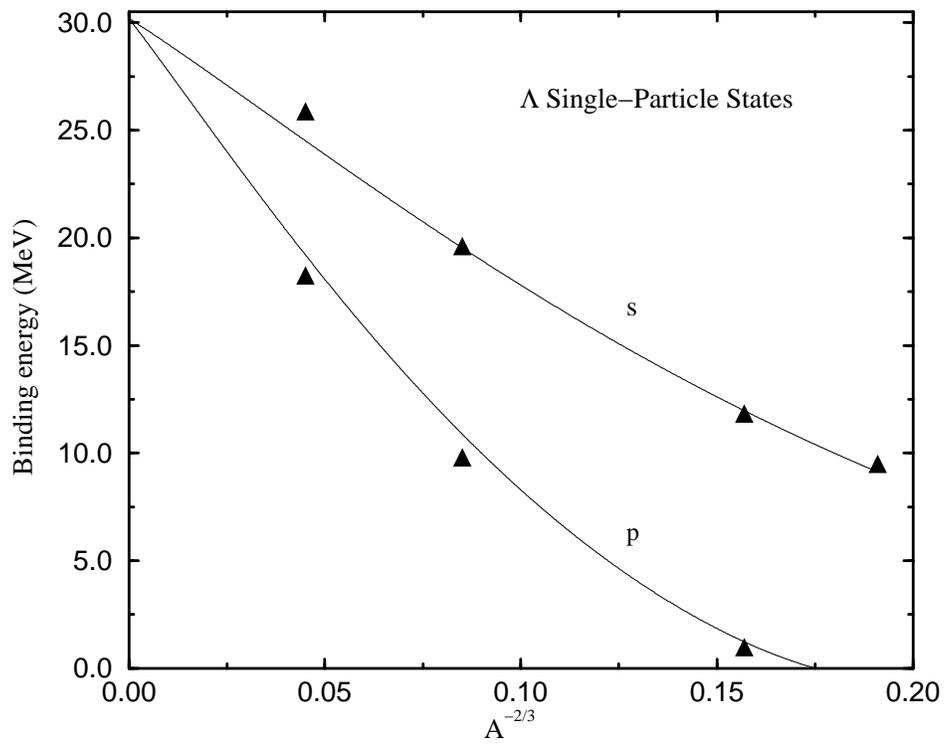}}
       \end{picture}
\caption{ Calculated $\Lambda$ binding energies in $1 s$ and $1 p$ single-partkce orbits for different
nuclei.  The curves correspond to the solutions obtained for a Woods-Saxon potential
whose parameters are defined in the text.}
   \label{fig:ener}
\end{figure}

Traditionally, the systematics of $\Lambda$ single-particle binding
energies
has been studied by using a phenomenological Woods-Saxon potential
\begin{equation}
      V_{WS}(r)=\frac{V_0}{1+\exp{[(r-R)/a]}},
      \label{eq:ws}
\end{equation}
with a fixed diffusivity $a$ and depth $V_0$,  and a radius $R=r_0
A^{1/3}$.
A good parameterization of the experimental data is obtained with
$V_0=-30.7$ MeV, $r_0=1.1$ fm and $a=0.6$ fm \cite{mille88}. 
 A more refined analysis of the same authors allows for a
smooth dependence of $r_0$ in $A$ and a slightly shallower
potential ($V_0=-28$ MeV) with a larger radius, $r_0(A)=(1.128 +
0.439 A^{-2/3}$) fm, provides a better agreement with the experimental $\Lambda$
binding energies. 
We have performed a similar analysis for  the calculated $\Lambda$
 binding energies. In principle, 
the calculated self--energy is non-local both in $k$-space and in
$r$-space.
However, in a previous work \cite{hjort96} we have shown that one can generate a
local representation of the self--energy by performing an appropriate
average of the non-local self--energy $\Sigma_{\alpha}(r,r')$, where
$\alpha$ indicates the quantum numbers of the single-particle state, over
the
coordinate $r'$. This local
representation might, in first approximation, be characterized by the
shape of a Woods-Saxon potential. Instead of doing this average, a possible 
strategy is to assume a Woods-Saxon shape, fix the depth and the
diffusivity independent of the mass number and determine the radius
$R$ by requiring the Woods-Saxon potential to reproduce the same
eigenvalue than the microscopic non-local energy dependent
self--energy. A reasonable value for the depth $V_0$ is the $\Lambda$
binding energy in nuclear matter, which is taken to be $-30.2$ MeV, and for
the diffusivity $a=0.6$ fm. The resulting values of $R$ when we apply this
procedure to the $s$ deepest state of $^{13}_{\Lambda}$C,
$^{17}_{\Lambda}$O and $^{41}_{\Lambda}$Ca are 2.25 fm, 2.53 fm and
3.82 fm respectively. Fitting these three values with a functional form similar
to the one used in Ref. \cite{mille88} for the analysis of the experimental data
one obtains $r_0(A)=(1.229 - 1.390 A^{-2/3})$ fm. In order to visualize the quality
of these fits, we show in Fig. \ref{fig:ener} the binding energies for the
$s$ and $p$ waves of $^{12}_{\Lambda}$C, $^{17}_{\Lambda}$O, $^{41}_{\Lambda}$Ca and
$^{91}_{\Lambda}$Zr calculated with our non-local self--energies
(triangles) together 
with the values obtained with a Woods-Saxon potential with 
the parameters just defined above (solid lines). 
As the spin-orbit splitting is so small we 
have reported the average value of the $p_{3/2}$ and $p_{1/2}$ 
energies obtained from the non-local
self--energies and have
not considered any spin-orbit term in the  adjusted Woods-Saxon potential.   
The results of $^{209}_{\Lambda}$Pb have not been included in the plot because,
as mentioned before, the $s$ wave binding energy was larger than the
binding energy in nuclear matter which we have taken as the depth of the 
Woods-Saxon potential. 
The calculated binding energies are well reproduced by the Woods-Saxon shape
and, as expected, both partial waves extrapolate to the binding
energy for nuclear matter.

Of course the binding energies are not enough to characterize the
single-particle states since potentials giving rise to the same binding
energies can generate substantial differences in the corresponding
wave functions. Therefore, in order to analyze the microscopically
calculated self--energy we should also study the single-particle wave
functions.

\begin{figure}
       \setlength{\unitlength}{1mm}
       \begin{picture}(100,180)
       \put(5,10){\epsfxsize=14cm \epsfbox{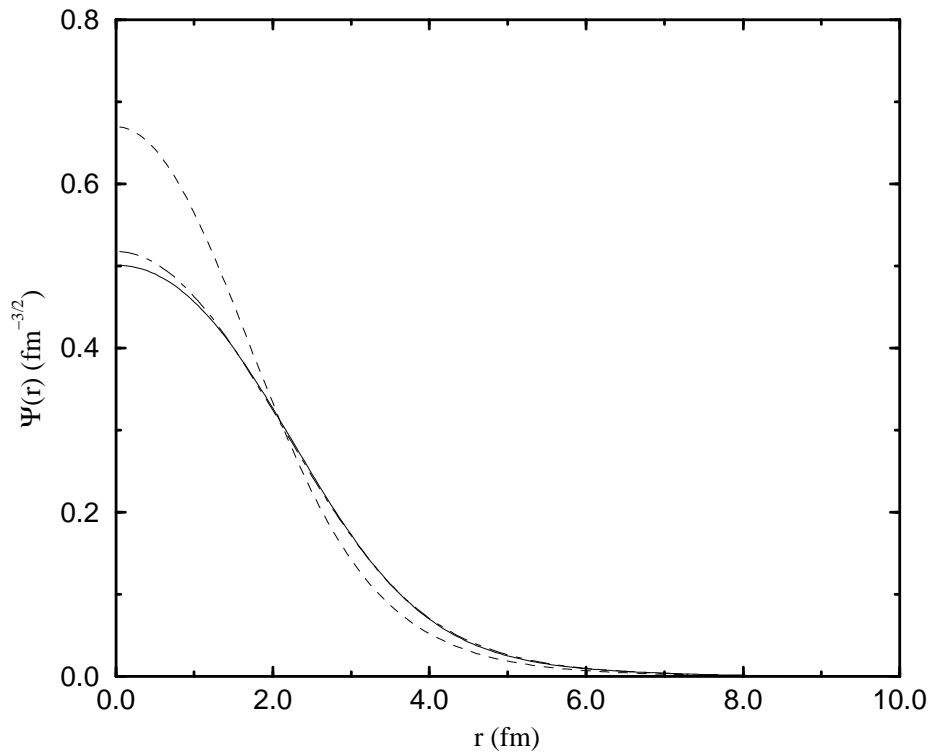}}
       \end{picture}
   \caption{The wave function in $r$-space for the $1s_{1/2}$ $\Lambda$ in 
            $^{17}_{\Lambda}$O obtained from the $\Lambda$ self--energy
(solid line)
            is compared with the ones obtained using a Woods-Saxon potential of fixed
            ($A$-independent) depth (dashed line) or with both radius and depth 
            adjusted (dot-dashed line) to maximize the overlap with the wave
            function provided by the self--energy.} 
     
   \label{fig:funcs}
\end{figure}

 To have a measure of the goodness of the wave
functions generated by the Woods-Saxon potential,  we calculate their 
overlap with the wave functions obtained by solving the
Schr\"odinger equation using the self--energy. The overlaps for
$^{13}_{\Lambda}$C, 
$^{17}_{\Lambda}$O, $^{41}_{\Lambda}$Ca and $^{91}_{\Lambda}$Zr are  0.9917,
0.9869, 0.9924 and 0.9853 respectively, which 
 are not close enough to 1 to guarantee
the equality of the wave functions. 
This is visualized in Fig. \ref{fig:funcs} , where the wave
function
for the $s$ wave in $^{17}_{\Lambda}$O obtained with the Woods-Saxon potential
(dashed-line) is compared with the one obtained directly from the
self--energy (solid line).

Another possibility would be to keep the diffusivity fixed and
adjust the radius $R$ and the depth $V_0$ to reproduce the eigenvalue and to
maximize
the overlap with the eigenfunction provided by the self--energy.  The
values of $V_0$ by applying this procedure to the $s$ wave of the three lighter
nuclei considered 
 above are respectively $-23.11$, $-23.56$ and $-27.84$ MeV
whereas the values of the radius $R$ are 2.92, 3.32 and 4.39 fm.  With these values
of $V_0$ and $R$ the overlaps are 0.9999 for the three nuclei
considered. The eigenfunction obtained by this procedure for
 the $^{17}_{\Lambda}$O is
also drawn in Fig.\ \ref{fig:funcs} (dot-dashed line) and shows a
large overlap with the self--energy eigenfunction (solid-line).

In conclusion, the single-particle energies of closed-shell nuclei
with one $\Lambda$ are well reproduced 
by using both the microscopic self--energy or the simpler
parametrization of
 Woods-Saxon type in the Schr\"odinger equation. However, the wave
functions provided by the microscopic self--energy differ from the ones
originated by a Woods-Saxon with a fixed depth and diffusivity and
a $A$-dependent radius. It is important to note that the mean square radius of the
self--energy eigenfunction is larger than that from the corresponding
Woods-Saxon wave function. This can have important consequences in 
the study the mesonic decay of these $\Lambda$ hypernuclei.
Indeed, it has been observed that the mesonic decay rates 
of light hypernuclei,
such as $^4_\Lambda$H, $^4_\Lambda$He and $^5_\Lambda$He, could be
better reproduced if the 
$\Lambda$ wave function was pushed out to the surface by the effect of
a repulsive hyperon-nucleus potential at short distances. This would   
favour the mesonic decay of these hypernuclei because the $\Lambda$ would
be exploring smaller nuclear density regions and the Pauli blocking effects,
which prevent the mesonic decay from occuring, would be less pronounced. 
The mesonic decay rates of light hypernuclei have been calculated using
repulsive 
$\Lambda$-nucleus potentials at short distances that have been obtained
either phenomenologically \cite{kuma95}, from a quark based bare $YN$
interaction \cite{straub93} or from a microscopic $YNG$ effective interaction 
folded with an extremely compact $^4$He density \cite{moto91}. At present,
no calculation exists that combines the use of an $YN$ effective interaction
with an appropiate density treatment of the host nucleus. Our method
provides such ingredients and has been shown to produce $\Lambda$ wave 
functions that are pushed out to the surface. This is a consequence of the
non-localities of the self--energy and might not be related to 
a repulsive character of the $\Lambda$-nucleus potential at short distances.
The implications of our
results on the mesonic decay of $\Lambda$ hypernuclei will be 
explored in a future work.

\section{Conclusions}
\label{conclusions}

We have analyzed a method to obtain the effective hyperon-nucleon
interaction in finite nuclei based on an expansion over a
$G$-matrix calculated in nuclear matter at fixed density and starting energy. 
The purpose of this study is to set up a reliable frame for
hypernuclear structure calculations with the aim of obtaining information about the
hyperon-nucleon interaction, complementary to that provided by  
hyperon-nucleon scattering experiments.
 
We have shown, by explicit calculation of the $\Lambda$ and $\Sigma$
single-particle energies in $^{17}_Y$O, that truncating the expansion
over the nuclear matter $G$-matrix at second order gives results that 
are very stable against variations of the density and starting energy
used in the $G$-matrix. Moreover, both first and second order terms depend
quite strongly on those parameters. This is an indication that
the density dependent effects considered when treating explicitly the
finite
size of the nucleus are very important and, therefore,  
they might not be well approximated by energy independent and local effective 
interactions
which start from a parametrized  
nuclear matter $G$-matrices evaluated at an average density. 
We note that the use of local and density averaged effective
interactions can be extremely useful in detecting similarities and differences
among the various hyperon-nucleon potentials. However, if the aim is 
to fine-tune the bare $YN$ interactions to reproduce the spectroscopic
data of
hypernuclei, an appropiate effective interaction for the 
hypernucleus under study, as the one provided by our method, is in order.
In particular, the two interactions employed in the present work give rise
to very attractive $\Sigma$ binding energies, while the $\Sigma^-$ atomic
data and
$(K^-,\pi^-)$ spectra seem to indicate $\Sigma$-nucleus potential depths
similar to that of the $\Lambda$ or even shallower. 
 
Although the method can be viewed as an alternative
way of building up a finite nucleus effective interaction, it
provides also the complete energy dependence of
the
hyperon self--energy. This allows in turn for a study of not only the
bound
states, as done here, but also the scattering states. This is especially  
of interest in the analysis of hypernuclear production
reactions which yield a large amount of quasifree hyperons.

We have obtained local Woods-Saxon $\Lambda$-nucleus potentials that
reproduce the $\Lambda$ single-particle energies of several hypernuclei.
However, the wave functions obtained from our non-local self--energy
are far more extended and can be simulated only when we
allow the Woods-Saxon potential  to have an $A$-dependent depth and a relatively 
larger radius. 
This can have important implications on the
weak mesonic decay of $\Lambda$ hypernuclei, which will be explored in
a future work.  

\section{Acknowledgements}
This work is partially supported by the DGICYT contract No. PB95-1249
(Spain) and by the Generalitat de Catalunya grant No. GRQ94-1022. One of
the authors (I.V.) wishes to acknowledge support from a doctoral fellowship 
of the Ministerio de Educaci\'on y Cultura (Spain).

\begin{table}               
\bigskip
\bigskip
\caption{Dependence of the $\Lambda$ single-particle energy in $^{17}_\Lambda$O
on the starting energy of the nuclear matter $G$-matrix. Our notation
is $\omega= <B_N> + B_\Lambda(k=0)$, with $<B_N>=-50$ MeV.
}
\bigskip
\bigskip
\begin{tabular}{c| cc |cc}
$k_F=1.36$ fm$^{-1}$ \phantom{caca}&
\multicolumn{2}{c|}{Nijmegen}
& \multicolumn{2}{c}{J\"ulich} \cr \hline
$\omega$ \phantom{cac}& $HF$ & $HF + 2p1h$ \phantom{ca}& $HF$ & $HF + 2p1h$ \cr
(MeV) \phantom{cac} & (MeV) & (MeV) \phantom{cac} & (MeV) & (MeV) \cr
\hline
$-100$ \phantom{cac}& $-3.83$ & $-7.43$ \phantom{cac}& $-9.25$ & $-11.85$ \cr
$-80$  \phantom{cac}& $-4.76$ & $-7.39$ \phantom{cac}& $-10.15$ & $-11.83$ \cr
$-50$  \phantom{cac}& $-5.59$ & $-7.36$ \phantom{cac}& $-11.73$ & $-11.84$ \cr
\end{tabular}
\label{tab:olambda}
\end{table}

\begin{table}
\caption{Dependence of the $\Sigma^0$ single-particle energy in
$^{17}_{\Sigma^0}$O
on the starting energy of the nuclear matter $G$-matrix. Our
notation
is $\omega= <B_N> + B_\Sigma(k=0) + \Delta$, with $<B_N>=-50$ MeV
and $\Delta=M_\Sigma-M_\Lambda$. }
\bigskip
\bigskip
\begin{tabular}{c| cc |cc}
$k_F=1.36$ fm$^{-1}$ \phantom{caca}& \multicolumn{2}{c|}{Nijmegen} &
\multicolumn{2}{c}{J\"ulich} \cr
 \hline
$\omega$ \phantom{cac}& $HF$ & $HF + 2p1h$ \phantom{ca} & $HF$ & $HF + 2p1h$ \cr
(MeV) \phantom{cac}& (MeV)  & (MeV) \phantom{cac}& (MeV) & (MeV) \cr
\hline
$  0$ \phantom{cac}& $-0.16$   & $-22.79$ \phantom{cac}& $-36.70$ & $-50.70$ \cr
$ 20$ \phantom{cac}& $-2.01$  & $-23.35$ \phantom{cac}& $-40.38$ & $-50.94$ \cr
$ 50$ \phantom{cac}& $-10.65$ & $-24.62$ \phantom{cac}& $-51.34$ & $-50.38$ \cr
\end{tabular}
\label{tab:osigma}
\end{table}

\begin{table}               
\caption{Dependence of the $\Lambda$ single-particle energy in $^{17}_\Lambda$O
on the Fermi momentum of the nuclear matter $G$-matrix. Our notation
is $\omega= <B_N> + B_\Lambda(k=0)$, with $<B_N>=-50$ MeV.
}
\bigskip
\bigskip
\begin{tabular}{c| cc |cc}
$\omega=-50$ MeV \phantom{caca}& \multicolumn{2}{c|}{Nijmegen} &
\multicolumn{2}{c}{J\"ulich} \cr \hline
$k_F$ \phantom{cac}& $HF$ & $HF + 2p1h$ \phantom{caca} & $HF$ & $HF + 2p1h$ \cr
(fm$^{-1}$) \phantom{cac}& (MeV)  & (MeV) \phantom{cac} & (MeV) & (MeV) \cr
\hline
$1.00$ \phantom{cac} & $-9.33$ & $-7.30$ \phantom{cac}& $-13.71$ & $-11.74$ \cr
$1.25$ \phantom{cac} & $-7.66$ & $-7.34$ \phantom{cac}& $-12.56$ & $-11.83$ \cr
$1.36$ \phantom{cac} & $-5.59$ & $-7.36$ \phantom{cac}& $-11.73$ & $-11.84$ \cr
\end{tabular}
\label{tab:dlambda}
\end{table}

\begin{table}
\caption{Dependence of the $\Sigma^0$ single-particle energy in
$^{17}_{\Sigma^0}$O
on the Fermi momentum of the nuclear matter $G$-matrix. Our
notation
is $\omega= <B_N> + B_\Sigma(k=0) + \Delta$, with $<B_N>=-50$ MeV
and $\Delta=M_\Sigma-M_\Lambda$. }
\bigskip
\bigskip
\begin{tabular}{c| cc |cc}
$\omega=0$ MeV \phantom{caca}& \multicolumn{2}{c|}{Nijmegen} &
\multicolumn{2}{c}{J\"ulich} \cr
 \hline
$k_F$ \phantom{cac}& $HF$ & $HF + 2p1h$  \phantom{caca}& $HF$ & $HF + 2p1h$ \cr
(fm$^{-1}$) \phantom{cac}& (MeV)  & (MeV) \phantom{cac}& (MeV) & (MeV) \cr
\hline
$1.00$ \phantom{cac} & $-2.64$  & $-25.03$  \phantom{cac}& $-43.94$ & $-51.74$ \cr
$1.25$ \phantom{cac} & $-0.82$  & $-23.37$  \phantom{cac}& $-39.41$ & $-51.16$ \cr
$1.36$ \phantom{cac} & $-0.16$  & $-22.79$  \phantom{cac}& $-36.70$ & $-50.70$ \cr
\end{tabular}
\label{tab:dsigma}
\end{table}
\vfil\eject
\begin{table}
\caption{ $\Lambda$ binding energies in the $1 s_{1/2}$, $1 p_{3/2}$ and
 $1 p_{1/2}$ single-particle orbits 
  for different
nuclei. The available experimental data, indicating the hypernucleus for which they have been
measured, are taken from the compilation of \protect\cite{bando90} supplemented by new measures reported
in \protect\cite{pile91} and \protect\cite{hase96}. All the results have been derived from the J\"ulich 
B interaction.}
\bigskip
\bigskip
\begin{tabular}{ c | c c c c  }
Hypernuclei \phantom{caca}& Orbit & $HF$ & $HF+2p1h$ &  Exp{}  \\
\hline
&&&&($^{13}_{\Lambda}$C) \\
$^{13}_{\Lambda}$C \phantom{cac}& $1 s_{1/2}$ & $-7.93$ & $-9.48$ & $-11.69$  {}\\
\hline
&&&&($^{16}_{\Lambda}$O){}\\
$^{17}_{\Lambda}$O \phantom{cac}& $1 s_{1/2}$ & $-10.15$ & $-11.83$ & $-12.5$ {}\\
 & $1 p_{3/2}$ &       & $-0.87$ & $-2.5$ ($1p$){}\\
 & $1 p_{1/2}$ & $-0.08$ & $-1.06$ & \\
\hline
&&&&($^{40}_{\Lambda}$Ca){}\\
$^{41}_{\Lambda}$Ca \phantom{cac}& $1 s_{1/2}$ & $-16.85$ & $-19.60$ & $-20.$ {}\\
 & $1 p_{3/2}$ & $-6.70$      & $-9.64$ & $-12.$ ($1p$){}\\
 & $1 p_{1/2}$ & $-6.92$ & $-9.92$ & {}\\
\hline
&&&&($^{89}_{\Lambda}$Zr) \\
$^{91}_{\Lambda}$Zr \phantom{cac}& $1 s_{1/2}$ & $-22.24$ & $-25.80$ & $-23.$ {}\\
 & $1 p_{3/2}$ & $-14.74$      & $-18.19$ & $-16.$ ($1p$){}\\
 & $1 p_{1/2}$ & $-14.86$ & $-18.30$ & {}\\
\hline
&&&&($^{208}_{\Lambda}$Pb) \\
$^{209}_{\Lambda}$Pb \phantom{cac}& $1 s_{1/2}$ & $-26.28$ & $-31.36$ & $-27.$ {}\\
 & $1 p_{3/2}$ & $-21.22$      & $-27.13$  & $-22.$ ($1p$) {}\\
 & $1 p_{1/2}$ & $-21.30$ & $-27.18$ &{}\\
\end{tabular}
\label{tab:ener}
\end{table}


\end{document}